\documentclass[aps, showpacs, showkeys,nofootinbib,floatfix]{revtex4}
\usepackage{amsfonts}

\usepackage{amssymb}
\usepackage{amsmath}
\usepackage{graphicx}
\usepackage{hyperref}

\begin{document}

\title{Dirac quasinormal modes of a Schwarzschild black hole surrounded by free static spherically symmetric quintessence}

\author{Yu Zhang}
\email{zhangyu128@student.dlut.edu.cn}
\author{Yuan-Xing Gui}
\email{guiyx@dlut.edu.cn}
\author{Fei Yu}
\affiliation{School of Physics and Optoelectronic Technology, Dalian
University of Technology, Dalian, 116024, P. R. China}

\begin{abstract}
We evaluate the quasinormal modes of massless Dirac perturbation in
a Schwarzschild black hole surrounded by the free static spherically
symmetric quintessence by using the third-order WKB approximation.
The result shows that due to the presence of quintessence, the
massless field damps more slowly. The real part of the quasinormal
modes increases and the the absolute value of the imaginary part
increases when the state parameter $w_q$ increases. In other words,
the massless Dirac field decays more rapidly for the larger $w_q$.
And the peak value of potential barrier gets higher as $|k|$
increases and the location of peak moves along the right for fixed
$w_q$.
\end{abstract}


\maketitle

\section {Introduction}
Quasinormal modes of black holes play a important role in the black
hole theory. They are regards as the \lq\lq characteristic
sound\rq\rq of black holes. The concept of QNMs was firstly pointed
out by Vishveshwara\cite{1} in calculations of the scattering of
gravitational waves by a black hole. In recent decades, many people
have done a lot of work of the quasinormal modes of black
holes\cite{2}-\cite{10}. They have investigate the perturbation such
as: gravitational perturbation\cite{11}, scalar
perturbation\cite{12}, electromagnetic perturbation\cite{13}, Dirac
perturbation\cite{14}-\cite{17}, and so on. There are many methods
to calculate the QNMs of the perturbation, for example, WKB
approximation\cite{18}-\cite{20}, the \lq\lq potential
fit\rq\rq\cite{21}, and the method of continued fractions\cite{22},
etc. In this paper, we use the third WKB approximation.
 \\There are three major motivations to study the QNMs in detail. First, QNMs may
provide valuable help to identify black hole
parameters\cite{23}\cite{24}. Second, an important aspect of QNMs
studies has been related to the AdS/CFT conjecture. By computing QN
frequencies in AdS spcectime, we can obtain a prediction for the
thermalization timescale in the strongly coupled CFT. It is argued
that string theory in AdS space is equivalent to CFT in one less
dimension\cite{25}. And the third is the relation between QNMs and
black hole area quantization\cite{26}-\cite{28}.
 \\ Recently, V. V. Kiselev\cite{29} has considered
Einstein's field equations for a black hole surrounded by the static
spherically symmetric quintessential matter and obtained a new
solution that depends on the state parameter $ w_{q}$ of the
quintessence. And Chen and Jing evaluated the quasinormal
frequencies of massless scalar field perturbation\cite{30} around
the black hole which is surrounded by quintessence. In other papers,
we have investigate the gravitational perturbation\cite{31} and the
electromagnetic perturbation\cite{32}. In this paper, we consider
the massless Dirac field in this situation.

\section{Dirac equation in the space time of schwarzschild black hole surrounded by quintessence}
The metric for the spacetime of Schwarzschild black hole surrounded
by the static spherically-symmetric quintessence is given
by\cite{30}
 \begin{equation}\label{eq:1}
 ds^{2}=-fdt^{2}+f^{-1}dr^{2}+r^{2}(d\theta^{2}+sin^{2}\theta
d\phi^{2}),\end{equation}
 with
 \begin{eqnarray}
 f=1-\frac{2M}{r}-\frac{c}{r^{3w_{q}+1}}
 \end{eqnarray}
where $M$ is the black hole mass, $w_{q}$ is the quintessential
state parameter, $c$ is the normalization factor related to
$\rho_{q}=-\frac{c}{2}\frac{3w_{q}}{r^{3(1+w_{q})}}$, and $\rho_{q}$
is the density of quitenssence.
\\ The massless Dirac equation in the black hole spacetime can be
written as\cite{33}
 \begin{eqnarray}
  [\gamma^a e_a^\mu(\partial _\mu+\Gamma_\mu)]\Psi=0, \label{Di}
\end{eqnarray}
where  $e_a^\mu$ is the inverse of the tetrad $e_\mu^a$ defined by
the metric $g_{\mu\nu}$,
\begin{eqnarray}
g_{\mu\nu}=\eta_{ab}{e_{\mu}}^{a}{e_{\nu}}^{b}
\end{eqnarray}
with $\eta_{ab}={\rm diag}(-1,1,1,1)$ being the Minkowski metric.
\\$\gamma^a$ is the Dirac matrix, and $\Gamma_\mu $ is the spin
connection which is given by
\begin{eqnarray}
 \Gamma_\mu=
\frac{1}{8}[\gamma^a,\gamma^b] e_a^\nu e_{b\nu;\mu}
\end{eqnarray}
where
$e_{b\nu;\mu}=\partial_{\mu}e_{b\nu}-\Gamma^{\alpha}_{\mu\nu}e_{b\alpha}$
is the covariant derivative of $e_{b\nu}$, and
$\Gamma^{\alpha}_{\mu\nu}$ is the Christoffel symbols.
\\we take the tetrad as
\begin{eqnarray}
    e_\mu^a=diag(\sqrt{f}, \frac{1}{\sqrt{f}}, r, r \sin \theta).
\end{eqnarray}
$\Gamma_{\mu}$ can be expressed as
\begin{eqnarray}
    \Gamma_{0}=\frac{1}{4}f'\gamma^{1}\gamma^{0}, \Gamma_{1}=0,
    \Gamma_{2}=\frac{1}{2}\sqrt{f}\gamma^{1}\gamma^{2},
    \Gamma_{3}=\frac{1}{2}(\sin\theta\sqrt{f}\gamma^{1}\gamma^{3}+\cos\theta\gamma^{2}\gamma^{3})
\end{eqnarray}
 Then, the Dirac equation (\ref{Di}) becomes
\begin{eqnarray}
    \frac{\gamma^{0}}{\sqrt{f}}\frac{\partial \Psi}{\partial t}+\sqrt{f}
    \gamma^{1} \left(\frac{\partial }{\partial r}+\frac{1}{r}+\frac{1}{4 f}
    \frac{d f}{d r} \right) \Psi+\frac{\gamma^{2}}{r}(\frac{\partial }
    {\partial \theta}+\frac{1}{2}cot\theta)\Psi+\frac{\gamma^{3}}{r
    \sin\theta}\frac{\partial \Psi}{\partial \varphi}=0. \label{Di1}
\end{eqnarray}
Define
\begin{eqnarray}
    \Psi=f^{-\frac{1}{4}}\Phi,
\end{eqnarray}
Eq. (\ref{Di1}) becomes
\begin{eqnarray}\label{Di2}
    \frac{\gamma^{0}}{\sqrt{f}}\frac{\partial \Phi}{\partial t}+\sqrt{f}
    \gamma^{1}\left(\frac{\partial }{\partial r}+\frac{1}{r} \right)
    \Phi+\frac{\gamma^{2}}{r}(\frac{\partial }{\partial \theta}+
    \frac{1}{2}cot\theta)\Phi+\frac{\gamma^{3}}{r \sin\theta}
    \frac{\partial \Phi}{\partial \varphi}=0. \label{Di2}
\end{eqnarray}
Define the tortoise coordinate $r_{*}$ as
\begin{eqnarray}
     r_{*}=\int\frac{dr}{f}
\end{eqnarray}
Introducing the ansatz
\begin{eqnarray}
    \Phi=\left(
\begin{array}{c}
\frac{i G^{(\pm)}(r)}{r}\phi^{\pm}_{jm}(\theta, \varphi) \\
\frac{F^{(\pm)}(r)}{r}\phi^{\mp}_{jm}(\theta, \varphi)
\end{array}\right)e^{-i \omega t},
\end{eqnarray}
with spinor angular harmonics
\begin{eqnarray}
    \phi^{+}_{jm}=\left(
\begin{array}{c}
\sqrt{\frac{l+\frac{1}{2}+m}{2 l+1}}Y^{m-1/2}_l \\
\sqrt{\frac{l+\frac{1}{2}-m}{2 l+1}}Y^{m+1/2}_l
\end{array}\right), \ \ \ \ \ \ \ \ \ \ \ \  (for \ \ j=l+\frac{1}{2}),
\nonumber
\end{eqnarray}
\begin{eqnarray}
    \phi^{-}_{jm}=\left(
\begin{array}{c}
\sqrt{\frac{l+\frac{1}{2}-m}{2 l-1}}Y^{m-1/2}_l \\
-\sqrt{\frac{l+\frac{1}{2}+m}{2 l-1}}Y^{m+1/2}_l
\end{array}\right), \ \ \ \ \ \ (for \ \ j=l-\frac{1}{2}), \nonumber
\end{eqnarray}
equations \ref{Di2} can be written in the form
\begin{eqnarray}
\left(\begin{array}{cc}0  & -\omega \\ \omega  & 0
\end{array}\right) \left(
\begin{array}{c}F^{\pm} \\G^{\pm}\end{array}\right)-\frac{\partial}
{\partial r_*}\left(\begin{array}{c}F^{\pm}
\\G^{\pm}\end{array}\right)+\sqrt{f}\left(\begin{array}{cc}
\frac{k_{\pm}}{r}  & 0 \\ 0 &  -\frac{k_{\pm}}{r} \end{array}\right)
\left(
\begin{array}{c}F^{\pm} \\G^{\pm}\end{array}\right)=0.
\end{eqnarray}
The cases for $(+)$ and $(-)$ in the functions $F^{\pm}$ and
$G^{\pm}$ can be put together giving
\begin{eqnarray}
    \frac{d^2 F}{d r_*^2}+(\omega^2-V_1)F&=&0, \label{even}\\
    \frac{d^2 G}{d r_*^2}+(\omega^2-V_2)G&=&0, \label{odd}
\end{eqnarray}
with
\begin{eqnarray}
V_1&=&\frac{\sqrt{f}|k|}{r^2}\left(|k|\sqrt{f}+\frac{r}{2}\frac{d f}
{d r}-f\right), \ \ \ \left(for \ \  k=j+\frac{1}{2},\ \ \ \ \ \  \
\ \ \  and \ \ j=l+\frac{1}{2}\right), \label{V1} \\
V_2&=&\frac{\sqrt{f}|k|}{r^2}\left(|k|\sqrt{f}-\frac{r}{2}\frac{d f}
{d r}+f\right), \ \ \ \left(for \ \  k=-\left(j+\frac{1}{2}\right),\
\  and \ \ j=l-\frac{1}{2}\right). \label{V2}
\end{eqnarray}
The potentials $V_1$ and $V_2$ are supersymmetric partners derived
from the same superpotential. We shall concentrate just on Eq.
\ref{even} with potential $V_1$ in evaluating the quasinormal mode
frequencies for the massless Dirac field by the third WKB
approximation in the following sections.

\section{Massless Dirac Quasinormal frequencies}
In this section we evaluate the QN frequencies for the massless
Dirac field in the space time of the Schwarzschild black hole
surrounded by quintessence using the third WKB approximation. WKB
method was firstly developed by B. F. Schutz and C. M. Will\cite{18}
at the second order. later, S. Iyer and C. M. Will\cite{19}
developed the method to the third order and R. A. Konoplya\cite{20}
extended it to the sixth order.The accuracy of the WKB formula is
better with a larger multipole number $l$ and a smaller overtone
$n$. The formula for the complex quasinormal frequencies
$\omega$(for the third order WKB method) is
\begin{equation}\label{11}
    \omega^{2}=\left [V_{0}+(-2V''_{0})^{1/2}\Lambda\right ]-i(n+\frac{1}{2})(-2V''_{0})^{1/2}(1+\Omega)
\end{equation}
where
\begin{align*}
    \Lambda=\frac{1}{(-2V''_{0})^{1/2}}\left\{\frac{1}{8}\left(\frac{V^{(4)}_{0}}{V''_{0}}\right)(\frac{1}{4}+\alpha^{2})-\frac{1}{288}\left(\frac{V'''_{0}}{V''_{0}}\right)^{2}(7+60\alpha^{2})\right\}
\end{align*}
\begin{align}
   \Omega=&\,\frac{1}{-2V''_{0}}\Big\{\frac{5}{6912}\left(\frac{V'''_{0}}{V''_{0}}\right)^{4}(77+188\alpha^{2})\nonumber\\
   &-\frac{1}{384}\left(\frac{V'''^{2}_{0}V^{(4)}_{0}}{V''^{3}_{0}}\right)(51+100\alpha^{2})+\frac{1}{2304}\left(\frac{V^{(4)}_{0}}{V''_{0}}\right)^{2}(67+68\alpha^{2})\nonumber\\
    &+\frac{1}{288}\left(\frac{V'''_{0}V^{(5)}_{0}}{V''^{2}_{0}}\right)(19+28\alpha^{2})-\frac{1}{288}\left(\frac{V^{(6)}_{0}}{V''_{0}}\right )(5+4\alpha^{2})\Big \}\label{12}
\end{align}
and
\begin{equation}\label{13}
    \alpha=n+\frac{1}{2},\> V^{(n)}_{0}=\frac{d^{n}V}{dr^{n}_{*}}\Big|_{r_{*}=r_{*}(r_{p})}
\end{equation}
 Take $M=1, c=0.001$ and $M=1, c=0$ for our calculation. And $c=0$
 means there is no quintessence. Using the third-order WKB approximation,
 we can get the solutions as the table $1$ and table $2$ show(for $0\leqslant n<k$), where $l$ is the angular harmonic index,
 $n$ is the overtone number, $\omega$ is the complex quasinormal frequencies, $w_{q}$ is the quintessential state
parameter.\\

 The effective potential $V(r,k)$, which depends on the absolute value of $k$ and $w_q$, is in the form of a barrier.
 In Fig. 1, it shows the variation of effective potential with $|k|$ when $w_q=-\frac{1}{2}$ and $c=0.001$. The variation of
 the effective potential with $r$ which is respective to the quintessential state $w_q$ parameter for fixed
$|k|=5$ and $c=0.001$ is shown in Fig. 2.\\
\begin{figure}
    \includegraphics[angle=0, width=0.6\textwidth]{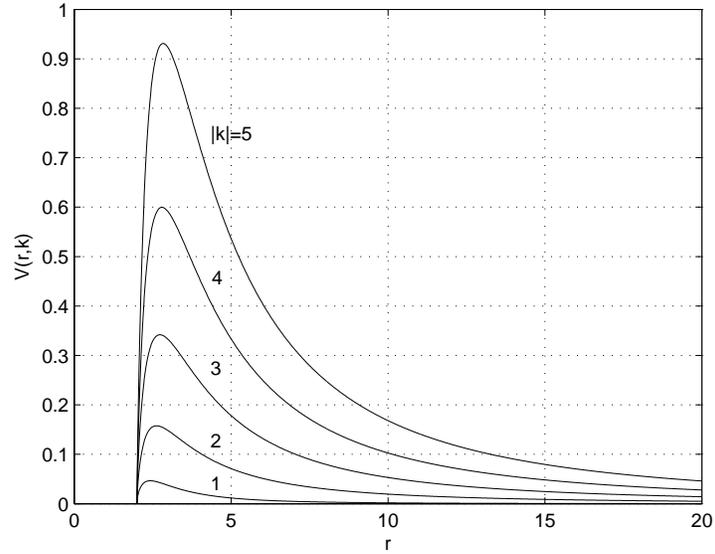}
\caption{Variation of the effective potential for the massless Dirac
field with $|k|$ in the Schwarzshild black hole surrounded by
quintessence for $w_{q}=-1/2$ and $c=0.001$.}
\end{figure}

\begin{figure}
    \includegraphics[angle=0, width=0.6\textwidth]{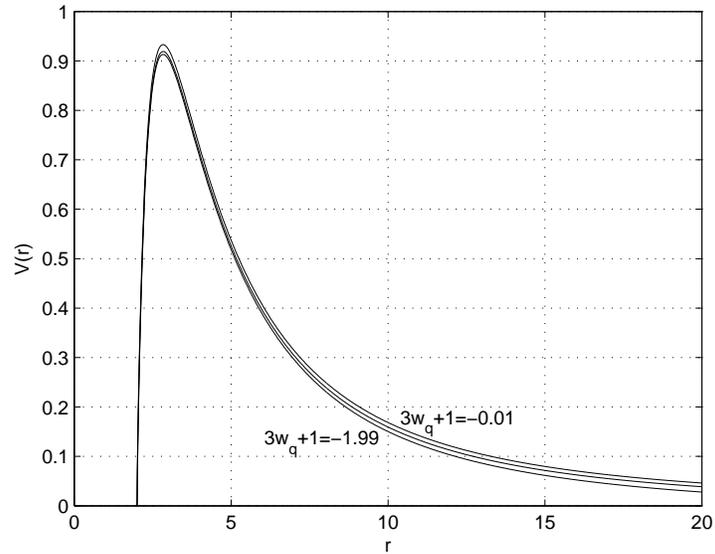}
\caption{Variation of the effective potential for the massless Dirac
field in the Schwarzshild black hole surrounded by quintessence with
$r$ for $|k|=5, c=0.001$ and $3w_{q}+1=-0.01, -1.7, -1.99$.}
\end{figure}

TABlE I:The quasinormal frequencies of massless Dirac perturbations
in the Schwarzshild black hole\cite{14} without quintessence(c=0).
\begin{equation}\label{121}
        \begin{tabular}{ccc|ccc}
    \hline\hline
    $|k|$ & $n$ & $\omega$&$|k|$ & $n$ & $\omega$ \\
    \hline
    $1$&$0$&$0.17645 -0.10011i$&$4$&$0$&$0.76719 -0.09628i$\\
    $$&$$&$$&$$&$1$&$0.75396 -0.29105i$\\
    $$&$$&$$&$$&$2$&$0.73045 -0.49088i$\\
    $$&$$&$$&$$&$3$&$0.69992-0.69571i$\\
    $2$&$0$&$0.37863 -0.09654i$&$5$&$0$&$0.96021 -0.09626i$\\
    $$&$1$&$0.35360 -0.29875i$&$$&$1$&$0.94959-0.29018i$\\
    $3$&$0$&$0.57369 -0.09632i$&$$&$2$&$0.92998 -0.48763i$\\
    $$&$1$&$0.55619 -0.29298i$&$$&$$3&$0.90358 -0.68924i$\\
    $$&$2$&$0.52729-0.49719i$&$$&$4$&$0.87205 -0.89441i$\\
    \hline\hline
  \end{tabular}\nonumber
\end{equation}

 TABlE II: The quasinormal frequencies of massless Dirac
perturbations in the black hole surrounded by quintessence for
$|k|=2$, $|k|=3$, $|k|=4$, $|k|=5$ and $c=0.001$.

 \begin{equation}\label{121}
 \nonumber
         \begin{tabular}{cc|ccc}
     \hline\hline
      $3w_{q}+1$ & $\omega(n=0)$ & $3w_{q}+1$ & $\omega(n=0)$ & $\omega(n=1)$ \\
     \hline
     $|k|=1$&$$&$|k|=2$&$$&$$\\
     $-0.01$&$0.17620-0.09990i$&$-0.01$&$0.37806-0.09635i$&$0.35310-0.29813i$\\
     $-0.4$&$0.17611-0.09977i$&$-0.4$&$0.37780-0.09623i$&$0.35287-0.29776i$\\
     $-0.8$&$0.17598-0.09956i$&$-0.8$&$0.37741-0.09605i$&$0.35252-0.29721i$\\
     $-1.2$&$0.17577-0.09926i$&$-1.2$&$0.37683-0.09579i$&$0.35199-0.29639i$\\
     $-1.6$&$0.17546-0.09881i$&$-1.6$&$0.37596-0.09541i$&$ 0.35118-0.29523i$\\
     $-1.99$&$0.17500-0.09819i$&$-1.99$&$0.37468-0.09489i$&$0.35001-0.29364i$\\
     \hline\hline
   \end{tabular}\nonumber
 \end{equation}

 \begin{equation}\label{546}
  \nonumber
    \begin{tabular}{cccccc}
     \hline\hline
     $3w_{q}+1$ & $\omega(n=0)$ & $\omega(n=1)$ & $\omega(n=2)$& $\omega(n=3)$ & $\omega(n=4)$ \\
     \hline
     $|k|=3$&$$&$$&$$&$$&$$\\
     $-0.01$&$0.57282-0.09613i$&$0.55536-0.29238i$&$0.52654-0.49617i$&$$&$$\\
     $-0.4$&$0.57240-0.09601i$&$0.55497-0.29203i$&$0.52618-0.49557i$&$$&$$\\
     $-0.8$&$0.57176-0.09584i$&$0.55436-0.29150i$&$0.52561-0.49466i$&$$&$$\\
     $-1.2$&$0.57079-0.09559i$&$0.55344-0.29073i$&$0.52476-0.49336i$&$$&$$\\
     $-1.6$&$0.56933-0.09524i$&$0.55205-0.28965i$&$0.52349-0.49152i$&$$&$$\\
     $-1.99$&$0.56718-0.09478i$&$0.55005-0.28822i$&$0.52169-0.48903i$&$$&$$\\
     $|k|=4$&$$&$$&$$&$$&$$\\
     $-0.01$&$0.76603-0.09608i$&$0.75283-0.29046i$&$0.72938-0.48988i$&$0.69892-0.69429i$&$$\\
     $-0.4$&$0.76546-0.09597i$&$0.75227-0.29011i$&$0.72885-0.48929i$&$0.69843-0.69345i$&$$\\
     $-0.8$&$0.76457-0.09579i$&$0.75141-0.28959i$&$0.72803-0.48841i$&$0.69765-0.69220i$&$$\\
     $-1.2$&$0.76321-0.09555i$&$0.75009-0.28884i$&$0.72678-0.48715i$&$0.69649 -0.69042i$&$$\\
     $-1.6$&$0.76116-0.09521i$&$0.74810-0.28782i$&$0.72491-0.48540 i$&$0.69476 -0.68791i$&$$\\
     $-1.99$&$0.75812 -0.09478i$&$0.74521-0.28649i$&$0.72224-0.48308i$&$0.69235 -0.68452i$&$$\\
     $|k|=5$&$$&$$&$$&$$&$$\\
     $-0.01$&$0.95876-0.09606i$&$0.94817-0.28959i$&$0.92860-0.48664i$&$0.90226-0.68783i$&$0.87081-0.89258i$\\
     $-0.4$&$0.95803-0.09595i$&$0.94745-0.28925i$&$0.92791-0.48606i$&$0.90160-0.68701i$&$0.87019-0.89151i$\\
     $-0.8$&$0.95689-0.09578i$&$0.94633-0.28873i$&$0.92682-0.48519i$&$0.90057-0.68578i$&$0.86920-0.88992i$\\
     $-1.2$&$0.95515-0.09554i$&$0.94462-0.28800i$&$0.92518-0.48396i$&$0.89899-0.68404i$&$0.86772-0.88766i$\\
     $-1.6$&$0.95250-0.09521i$&$0.94203-0.28701i$&$0.92270-0.48227i$&$0.89666-0.68162i$&$0.86554-0.88449i$\\
     $-1.99$&$0.94856-0.09480i$&$0.93823-0.28575i$&$0.91913-0.48009i$&$0.89336-0.67843i$&$0.86253-0.88020i$\\
    \hline\hline
  \end{tabular}\nonumber
\end{equation}

    \begin{figure}
    \includegraphics[angle=0, width=0.6\textwidth]{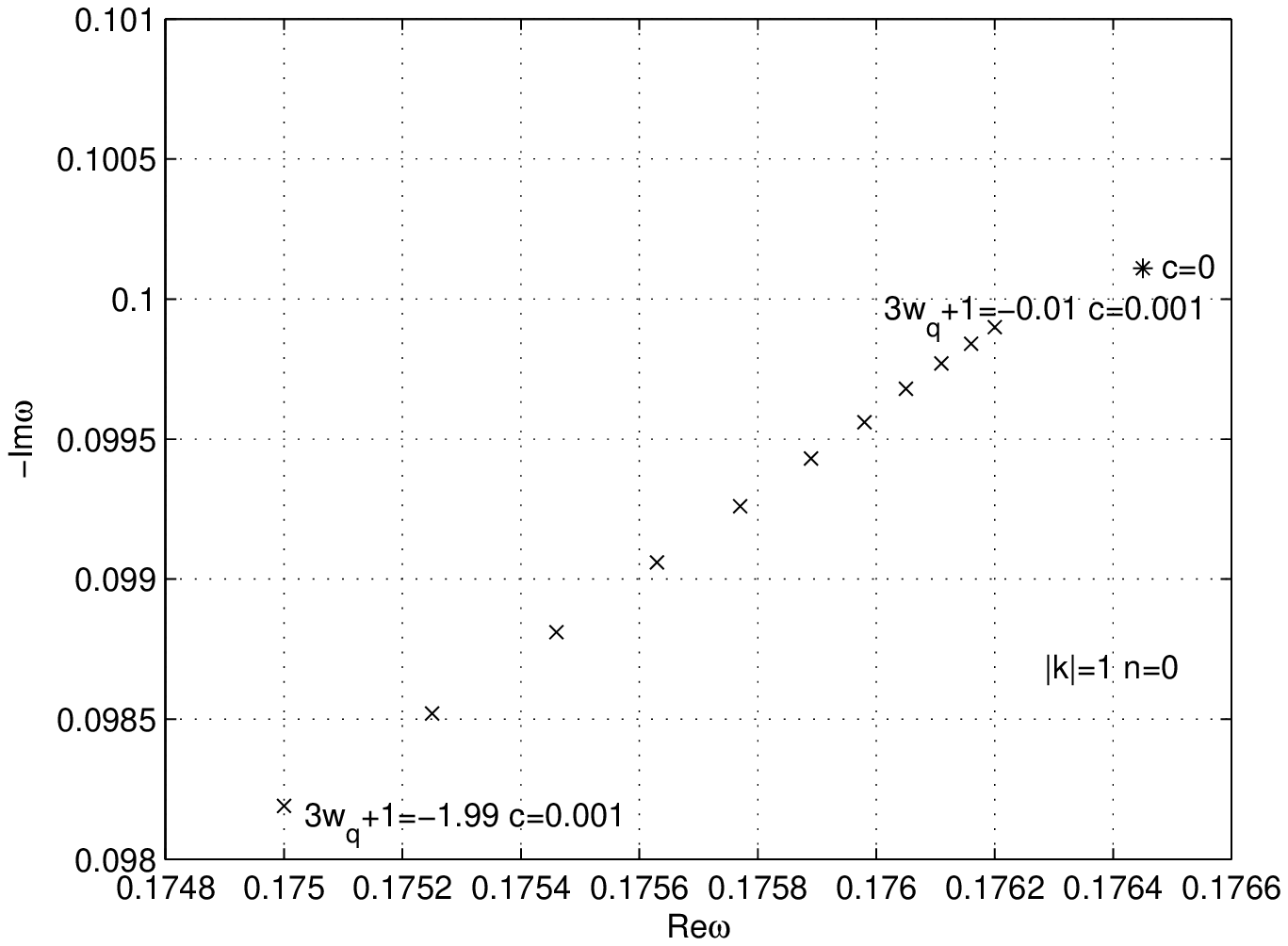}
    \includegraphics[angle=0, width=0.6\textwidth]{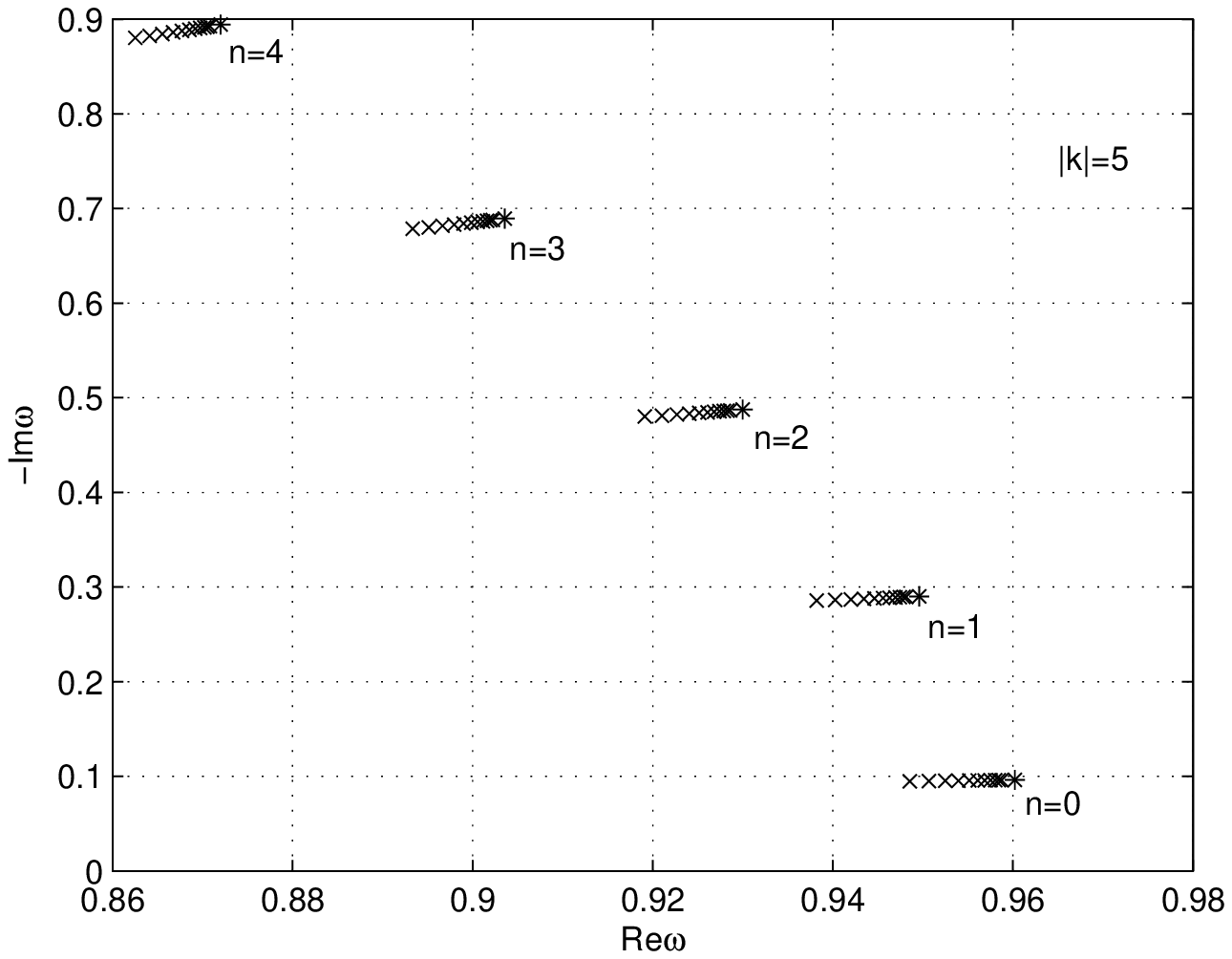}
\caption{The relationship between the real and imaginary
 parts of quasinormal frequencies for the massless Dirac
 perturbations in the background of the black hole surrounded by quintessence for
 fixed $c=0.001$ and no quintessence for $c=0$. $\times$ and
 $\ast$, respectively, refering to the value with quintessence and without quintessence.}
\end{figure}

\section{Discussion and Conclusion}

The quasinormal modes of a black hole present complex frequencies,
the real part of which represents the actual frequencies of the
oscillation and the imaginary part represents the damping.
\\ From Fig.1 we see that the peak value of potential barrier gets higher as
$|k|$ increases and the location of peak moves along the right for
fixed $w_q$. From Fig. 2, it shows that when the absolute $w_q$
increases, the peak value of potential barrier gets lower. Using the
third WKB approximation, we calculate the lowly decaying modes
frequencies for $k=1,2,3,4,5$. The complex quasinormal mode
frequencies for the massless Dirac field are listed in Table
I(without quintessence) and Table II(exist quintessence, and
c=0.001). We plot the relationship between the real and imaginary
parts of quasinormal frequencies with the variation of $w_{q}$(for
fixed $c=0.001$), compared with the situation without quintessence.
From Fig.3 we can find that for fixed $c$(unequal to 0) and $l$ the
absolute values of the real and imaginary parts decrease as the
absolute value of the quintessence state parameter $w_{q}$
increases. It means that when the absolute value of $w_{q}$ is
bigger, the oscillations damp more slowly. And the absolute values
of the real and imaginary parts of quasinormal modes with
quintessence are smaller compared with those with no quintessence
for given $l$ and $n$. That is to say, due to the presence of
quintessence, the oscillations of the massless Dirac fields damp
more slowly.

 \setcounter{secnumdepth}{-1}

\acknowledgements{Yu Zhang wishes to thank Ph.D LiXin Xu for his
helpful discussions. This work is supported by the National Natural
Science Foundation of China under Grant No. 10573004.}

\end{document}